\documentclass[12pt,journal,draftcls,letterpaper,onecolumn]{IEEEtran}

\usepackage{multirow} 
\usepackage{graphicx}
\usepackage[table]{xcolor} 
\usepackage{float}
\usepackage{mathtools}
\usepackage{upgreek} 
\usepackage{amsmath}  
\usepackage{amssymb}  
\usepackage{epstopdf}  
\usepackage{breqn}
\usepackage{fixltx2e} 
\usepackage{float}
\usepackage{morefloats}
\usepackage{setspace} 
\usepackage[mathscr]{euscript} 
\usepackage{enumitem,kantlipsum}
\usepackage{algorithm} 
\usepackage{algorithmic}

\newcommand{\comments}[1]{} 
\newcommand{\figref}[1]{Fig. \ref{#1}}
\usepackage{courier}
\usepackage{cite}

\begin{document}
\IEEEoverridecommandlockouts
  
\title{Relay Selection \& Power Allocation \\ for Maximizing Sum-Throughput of a \\ Buffered Relay Network }
\comments{
\author{\IEEEauthorblockN{Author 1, Author 2, Author 3 and Author 4}
\thanks{Author 1  is with Affiliation 1. Email: author1@affiliation1.com. Author 2  is with Affiliation 2. Email: author2@affiliation2.com. Author 3 
is with Affiliation 3. Email: author3@affiliation3.com. Author 4 is with Affiliation 4. Email: author4@affiliation4.com. The corresponding author is
Author 1.} } 
}

\author{\IEEEauthorblockN{Fahd Ahmed Khan, Zafar Abbas Malik, Ali Arshad Nasir and Mudassir Masood} 

\thanks{Fahd Ahmed Khan and Zafar Abbas Malik are with the School of Electrical Engineering and Computer Science (SEECS), National University of
Sciences and Technology (NUST), Islamabad, Pakistan. Email: \{fahd.ahmed, 14mseezmalik\}@seecs.edu.pk. Ali Arshad Nasir and Mudassir Masood are with the King Fahd University of Petroleum and Minerals (KFUPM), Dharan, Saudi Arabia. Email: \{anasir,mudassir\}@kfupm.edu.sa. The corresponding author is Fahd Ahmed Khan.}}
 
\maketitle     

\begin{abstract}
Considering two-hop cooperative communication with buffered half-duplex relays, this work jointly optimizes relay selection and transmit power at the relay to maximize the sum-throughput of the network subject to a minimum throughput requirement of each hop and transmit power constraint. The optimization problem is solved and two new buffer aware selection schemes, namely joint relay power allocation and selection scheme (JPASS) and ratio selection scheme (RSS) are proposed. Numerical results show that the average sum-throughput of both the proposed schemes is up to 40\% higher compared to the existing schemes in the literature.
\end{abstract} 
 


\section{Introduction}
Cooperative relaying assists the direct data transmission between source and destination and provides multiple benefits such as improved reliability
and coverage \cite{BibPabst2004,BibLaneman2004, BibErkip2003} . However, traditional half-duplex cooperative relaying requires at least two time-slots for the data transmission from the source to the destination \cite{BibLaneman2004,BibErkip2003, BibFahd2012,BibDing2012}. Various non-orthogonal protocols have been proposed to address this transmission-delay and improve the throughput (see \cite{BibDing2012} and references therein). One possible solution is to utilize buffers at the relays \cite{BibXia2008,BibIkhlef2012, BibZlatanov2014, BibNomikos2016}.

The provision of buffers at the relays can enhance the effective throughput by enabling source-to-destination transmission in a single time slot. This is achieved by selecting different relays for reception (\emph{receiving-relay}) and transmission (\emph{transmitting-relay}) in each time slot. One of the relay, i.e., the receiving-relay, receives data from the source and stores it in its buffer (that can be forwarded to the destination during another time slot) while the other selected relay, the transmitting-relay, transmits the data stored in its buffer to the destination \cite{BibXia2008,BibIkhlef2012, BibZlatanov2014, BibNomikos2016}.

An immediate question arises that how are the relays selected? Many different criterion have been proposed for selecting the relays
\cite{BibIkhlef2012,BibNomikos2014,BibNomikos2015,BibSimoni2016, BibCharalambous2019, BibKim2016}. In max-max relay selection (MMRS)
\cite{BibIkhlef2012}, a relay with the best source-relay (S-R) channel (having the highest signal-to-noise ratio (SNR)) and another relay with the
best relay-destination (R-D) channel are selected for reception and transmission, respectively, in the same time slot. Since the source and relay
nodes transmit during the same time-slot, the receiving-relay experiences interference from the transmitting-relay, which was not taken into account
in the selection criteria of \cite{BibIkhlef2012}. In \cite{BibNomikos2014}, this interference was taken into account and the authors proposed a
buffer-aided successive opportunistic relay (BA-SOR) selection scheme. {In BA-SOR, the selection of the S-R link was based on the channel-power-to-interference-channel-power ratio and max-min selection criteria was employed. A ``min-power" relay selection policy was proposed in
\cite{BibNomikos2015}, where the authors proposed to select the relay pair which achieves a pre-determined fixed throughput with minimum power. Even if the channel can support a higher throughput, using the proposed allocation in \cite{BibNomikos2015}, the throughput is limited to the pre-determined value. Considering a diamond relay network with only two relays, Simoni et al. in \cite{BibSimoni2016}, proposed a transmission mode selection policy to maximize the throughput of a the network.
 
In \cite{BibCharalambous2019}, a joint precoding matrix design at the source and relay-pair selection is proposed to maximize the signal-to-noise-and-interference ratio (SINR). In \cite{BibKim2016}, considering relays with multiple antennas, the authors proposed joint relay selection and beamforming design, while taking inter-relay interference (IRI) into consideration, to maximize the sum-throughput. The beamforming design directs the power from the transmitting relay towards the destination and limits the interference power to the receiving relay. 
The use of multiple antennas at the transmitter and/or the relays, as in \cite{BibCharalambous2019} and \cite{BibKim2016}, enables information beamforming towards the desired node and interference avoidance at the receiving-relay. However, internet-of-things (IoT)-related applications motivate the need of simple relays employing single-antenna transceivers, which don't have the freedom to take benefit from beamforming or precoding to tackle IRI.

In \cite{BibIkhlef2012,BibNomikos2014,BibNomikos2015,BibSimoni2016}, a parameter which is not optimized, is the power of the transmitting-relay. The power of the transmitting-relay can be adjusted to control the IRI and also improve the sum-throughput. The high power of the transmitting-relay results in large interference at the receiving-relay which results in lower SINR and thus, lower throughput of the S-R link. On the other hand, the lower power of the transmitting-relay results in lower SINR and thus lower throughput of the R-D link. In addition, unlike the min-power scheme in \cite{BibNomikos2015}, where the system throughput was limited, the power of the transmitting-relay can be optimized to maximize the sum-throughput. In this regard, we propose two novel selection schemes, namely, 1) \emph{Joint relay power allocation and selection scheme} (JPASS) and 2) \emph{Ratio selection scheme} (RSS) to optimize the relay-pair selection and power of the transmitting-relay that
maximizes the network sum-throughput (of S-R and R-D channels) while satisfying the power and threshold SINR constraints. Numerical results show that the achievable throughput of both the proposed schemes is higher compared to the existing schemes in the literature and when the relays have sufficiently large buffers, the JPASS yields the highest average sum-throughput and serves as a benchmark.
}
\comments{

Rest of this paper is organized as follows. Section II describes the system model and the problem formulation. Section III elaborates
the proposed relay selection schemes. The performance of the proposed relay selection scheme is analyzed by numerical simulations and the results and
discussion is presented in Section IV. The conclusions are presented in Section V.
}

\begin{figure}            
\centering
  \includegraphics[width=0.7\columnwidth]{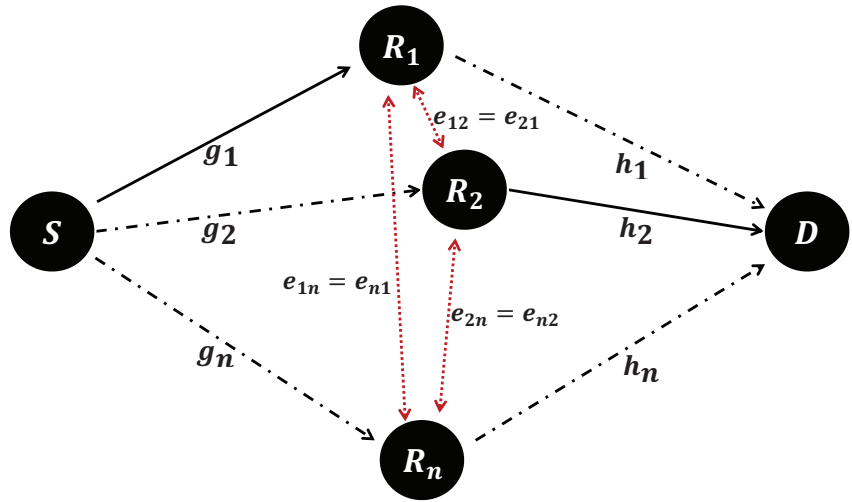}
  \caption{System Model}
   \label{FigSystem}
\end{figure}

\section{System Model and Problem Formulation}
Consider a cooperative network show in \figref{FigSystem} where a source node communicates to a destination node via $n$ decode-and-forward (DF) buffered relays. {The relays are assumed to be half-duplex (HD) relays with finite buffers.} We assume that the destination is out-of-reach from the source and there is no direct communication link between them. As the relays have buffers, one of the relay receives and stores the data in its buffer while another relay transmits the data already stored in its buffer during the same time slot. Thus, effectively, communication from the source-to-destination occurs in a single time slot instead of two time slots (required for conventional HD relaying), resulting in an increased
throughput. However, as two relays are communicating during the same time slot, the transmitting-relay causes interference to the receiving-relay.
\comments{Taking into account this interference, we perform joint relay pair selection and power optimization to maximize the sum-throughput of the network.}

The channel gain from the source to the $k$-th relay is denoted as $g_{k}$ and the channel gain from the $k$-th relay to the destination is denoted
as $h_k$. Inter-relay channel between the $i$-th and the $j$-th relay is denoted by $e_{ij}$. The channel is assumed to be reciprocal and have independent and identically distributed Rayleigh fading. Therefore, the channel powers, $|g_k|^2, |h_k|^2$ and $|e_{ij}|^2$ are exponentially distributed. The channel coherence time is assumed to be nearly equal to the total transmission time from source-to-destination. Thus, each transmission experiences uncorrelated fading, which will randomize the selected relays from one transmission to
another.\footnote{It is assumed that perfect global channel state information (CSI) is available. This ideal scenario will provide a benchmark on the performance of a buffered relay network. Discussion on CSI acquisition is omitted here due to space limitation. One can refer to
\cite{BibFahdDCE2011,BibIkhlef2012,BibQasim2014}, and the references therein, for a discussion on protocol for CSI acquisition. Deterioration in system performance due to imperfect CSI is discussed in Section IV.}$P_s$ denotes the transmitted power of source node and $P_j$ denotes the transmit power of the $j$-th transmitting-relay. The noise power at each node in the network is denoted by $N_0$.

\emph{Problem Formulation:} 
 Assuming that the $i$-th relay is selected for reception and the $j$-th relay is selected for transmission, the  SINR at the receiving-relay is given by
$\gamma_{ij} = \frac{P_s|g_i|^2 }{N_0+P_j |e_{ij}|^2 }$ while SNR at the destination is given by $\gamma_j = \frac{ P_j |h_j|^2 }{N_0}$. The second term in the denominator of $\gamma_{ij}$ is the interfering signal power from the transmitting-relay. Thus, large values of the channel gain $e_{ij}$ will cause high interference and low SINR at the receiving-relay. Therefore, relay pair selection must take into account the inter-relay channel $e_{ij}$. In addition, the transmitting-relay's power, $P_j$, also needs to be optimized so that it is sufficiently high to provide good SNR at the destination and low interference at the receiving-relay.

Taking into account this trade-off, we formulate and solve a joint optimization problem which selects a relay pair for transmission and reception and optimizes the relay transmission power, $P_j$, to maximize the instantaneous sum-throughput. The optimization problem is expressed as
\begin{equation}
\begin{aligned}
& \max_{P_j,i \neq j}
& & \uptau_{i,j}(P_j) = \mathcal{R}_{ij}^{SR}\left(P_j\right)+\mathcal{R}_j^{RD}\left(P_j\right)\\
&\text{subject to} 
& & \frac{ P_j |h_j|^2 }{N_0}  \geq \phi_{2}\\
& & & \frac{P_s|g_i|^2 }{N_0+P_j |e_{ij}|^2 } \geq \phi_{1}\\
& & & P_j \leq P_{max}\\
& & & \mathcal{R}^{RD}_j\left(P_j\right) t \leq Q_j \\
& & & Q_i +\mathcal{R}^{SR}_{ij}\left(P_j\right) t \leq Q_{max}
\label{eq:sum_throughput}
\end{aligned}
\end{equation}
where $\mathcal{R}_{ij}^{SR}\left(P_j\right)=\log_2\left(1+\frac{P_s|g_i|^2 }{N_0+P_j |e_{ij}|^2 }\right)$,
$\mathcal{R}_j^{RD}\left(P_j\right)=\log_2\left(1+\frac{ P_j |h_j|^2 }{N_0}\right)$, $i,j \ \in \{ 1,\hdots,n\}$ and {\bf $i\neq j$}, $P_{max}$ is the
maximum transmit power at the relay, $\phi_i=2^{{R}_i}-1$, ${R}_1$ and ${R}_2$ denotes the minimum rate requirement for the S-R and R-D link, $Q_{max}$ is the maximum buffer size, $Q_i$ denotes the number of bits stored in the buffer of relay $i$ and $t$ denotes the transmission slot duration. In \eqref{eq:sum_throughput}, the first two constraints guarantee a minimal quality of service (QoS) at the receiving-relay and the destination while the third constraint is the power constraint at the transmitting node. The fourth constraint ensures that the transmitting relay can transmit at most $Q_j$ bits stored in its buffer. The final constraint ensures that the receiving relay does not receive bits greater than the space available in its buffer.

\section{Relay Selection Schemes}

The optimization problem in \eqref{eq:sum_throughput} can be reformulated as
\begin{equation}
\begin{aligned}
& \max_{P_j,i \neq j}
& & \uptau_{i,j}(P_j) = \mathcal{R}_{ij}^{SR}\left(P_j\right)+\mathcal{R}_j^{RD}\left(P_j\right)\\
&\text{subject to} 
& & P_j^{min} \leq P_{j} \leq  P_{i,j}^{max}
\label{eq:sum_throughput2}
\end{aligned}
\end{equation}
where $P_j^{min}=\max \left\{
\frac{\phi_{2}N_{0}}{\left|h_{j}\right|^{2}}, \frac{P_s|g_i|^2 }{\left(2^{\left(Q_{max}/t-Q_i/t\right)}-1\right)|e_{ij}|^2 }-\frac{N_0 }{|e_{ij}|^2 } \right\}$ and $
P_{i,j}^{max}=\min\left\{P_{max},\frac{P_{s}\left|g_{i}\right|^{2}-\phi_{1}N_{0}}{\phi_{1}\left|e_{ij}\right|^{2}},
\frac{N_0\left(2^{Q_j/t}-1\right)}{|h_j|^2}\right\}$.

\emph{\bf Feasible range of $P_j$:} From the constraints in \eqref{eq:sum_throughput2}, if $P_j^{min}<P_{i,j}^{max}$, then the feasible range of
values for $P_{j}$ is
\begin{equation}
\label{eq:rangePr}
\mathfrak{R}=
\left[ P_j^{min}, P_{i,j}^{max} \right]
\end{equation}
If $P_j^{min}>P_{i,j}^{max}$, then $\mathfrak{R}=\phi$.

\comments{
\begin{equation}
\mathcal{R}=\begin{cases}
\left[ \phi_{2}\frac{N_{0}}{\left|h_{i}\right|^{2}},
\min\left\{ P_{max},\,\frac{P_{s}\left|g_{i}\right|^{2}-\phi_{1}N_{0}}{\phi_{1}\left|e_{ij}\right|^{2}}\right\} \right]  &
\phi_{2}\frac{N_{0}}{\left|h_{i}\right|^{2}}<\min\left\{P_{max},\,\frac{P_{s}\left|g_{i}\right|^{2}-\phi_{1}N_{0}}{\phi_{1}\left|e_{ij}\right|^{2}}\right\} \\
\hspace{40pt}\phi  & \phi_{2}\frac{N_{0}}{\left|h_{i}\right|^{2}}>\min\left\{
P_{max},\,\frac{P_{s}\left|g_{i}\right|^{2}-\phi_{1}N_{0}}{\phi_{1}\left|e_{ij}\right|^{2}}\right\}
\end{cases}
\end{equation}

$\mathcal{R}=\left[\phi_{2}\frac{N_{0}}{\left|h_{i}\right|^{2}}, \min\left\{
P_{max},\,\frac{P_{s}\left|g_{i}\right|^{2}-\phi_{1}N_{0}}{\phi_{1}\left|e_{ij}\right|^{2}}\right\}\right]$ given that condition
$\mathcal{C}=\phi_{2}\frac{N_{0}}{\left|h_{i}\right|^{2}}>\min\left\{
P_{max},\,\frac{P_{s}\left|g_{i}\right|^{2}-\phi_{1}N_{0}}{\phi_{1}\left|e_{ij}\right|^{2}}\right\} $ is true. If the channel gains and system
parameters are such that $\mathcal{C}$ is false, then no feasible $P_{j}$ is possible and relays have to operate in conventional HD mode.
}

\emph{\bf Optimal value of $P_{j}$:}
The objective function in \eqref{eq:sum_throughput2} can be expressed as $\log_2\left(\mathcal{C}(P_{j})\right)$, where
$\mathcal{C}(P_{j})=\frac{\left(1+\alpha_{1}\right)+\left(\alpha_{2}+\alpha_{3}+\alpha_{1}\alpha_{3}\right)P_{j}+\alpha_{2}\alpha_{3}P_{j}^{2}}{P_{j}\alpha_{2}+1}$,
 $\alpha_{1}=\frac{P_{s}}{N_{0}}\left|g_{i}\right|^{2}$, $\alpha_{2}=\frac{\left|e_{ij}\right|^{2}}{N_{0}}$ and
$\alpha_{3}=\frac{\left|h_{j}\right|^{2}}{N_{0}}$.
$\mathcal{C}(P_{j})$ is a quadratic over linear function, which is a convex function \cite{BibBoyd2004}. As, logarithm is a monotonically increasing
function, $\log_2\left(\mathcal{C}(P_{j})\right)$ will have maximum at the boundary of the function i.e. at either the lowest value of $P_j$ or the
maximum value of $P_j$.
From \eqref{eq:rangePr}, the lowest value of $P_j$ is $P_j^{min}$ and the largest value of $P_{j}$ is $P_{i,j}^{max}$. Based on this, the optimization problem in \eqref{eq:sum_throughput2}, can
be simplified as \comments{
\begin{equation}
\label{eq:optimalsolution}
\scriptsize
\begin{split}
\left\{P_j^*,i^*,j^*\right\}=
\arg \max_{P_{j},i,j}\left\{ \max \left\{\uptau_{i,j}\left(P_j^{min}\right),\uptau_{i,j}\left(P_{i,j}^{max}\right)\right\}\right\}
\end{split}
\end{equation}
}
\begin{equation}
\label{eq:optimalsolution}
\begin{split}
\left\{P_j^*,i^*,j^*\right\}=
\arg \max_{i \neq j,P_{j}\in \left\{P_j^{min}, P_{i,j}^{max}\right\}}  \uptau_{i,j}\left(P_j\right)
\end{split}
\end{equation}
Based on \eqref{eq:optimalsolution}, we first propose a joint relay power allocation and selection scheme (JPASS).  \comments{The
objective function is required to be evaluated at both extreme values of $P_{j}\,\,\forall i,j$ and the one resulting in higher objective function
will correspond to the optimal $P_j$ i.e.
} 
 
\subsection{\bf Joint Relay Power Allocation and Selection Scheme}

From \eqref{eq:optimalsolution}, it can be noted that the maximum sum-throughput value will occur only at either the minimum value of power,
$P_j^{min}$, or the maximum value, $P_{i,j}^{max}$. We take this into consideration and calculate the sum-throughput,
$\uptau_{i,j}(P_j)$, of all feasible links (which satisfy the constraint in \eqref{eq:rangePr} i.e. for which $P_j^{min}<P_{i,j}^{max}$). The
sum-throughput of all possible feasible links is evaluated only at the boundary points $P_j^{min}$ and $P_{i,j}^{max}$. The link pair (S-R link and R-D link)
which yields the highest sum-throughput is selected for transmission. The main steps for JPASS are as follows:
{
\begin{enumerate}
\item Find the set of feasible relays, $\mathfrak{F_{R}}$, for which $P_j^{min}<P_{i,j}^{max}$.
\item If $\mathfrak{F_{R}}=\phi$, i.e. no relay pair satisfies the constraints in \eqref{eq:sum_throughput2}, the network has to operate in
conventional HD mode. In this time slot, only S-R or R-D communication occurs. In order to maximize the throughput in this conventional HD mode, the link which yields the highest throughput is selected. Evaluate $\hat{i}=\arg
\max_{i, \left(Q(i)+\mathcal{R}_{ij}^{SR}\left(0\right)t\right)<Q_{max}} \mathcal{R}_{ij}^{SR}\left(0\right)$ and $\hat{j}=\arg
\max_{j,P_j<P_{max},\mathcal{R}_j^{RD}\left(P_j\right)t<Q_{j} }\mathcal{R}_j^{RD}\left(P_{j}\right)$. Select relay $\hat{i}$ for reception if
$\mathcal{R}_{\hat{i}j}^{SR}\left(0\right) > \mathcal{R}_{\hat{j}}^{RD}\left(P_{\hat{j}}\right)$ otherwise select relay $\hat{j}$ for transmission.
\comments{
\begin{equation}
\scriptsize
\begin{aligned}
& \text{Select relay $\hat{i}$ for reception if} & & \mathcal{R}_{\hat{i}j}^{SR}\left(0\right) > \mathcal{R}_{\hat{j}}^{RD}\left(P_{max}\right) \\
&\text{Select relay $\hat{j}$ for transmission if} & & \mathcal{R}_{\hat{i}j}^{SR}\left(0\right) <
\mathcal{R}_{\hat{j}}^{RD}\left(P_{max}\right) \\\\
\label{eq:HDmodeexpression}
\end{aligned}
\end{equation}
}
\comments{

\begin{equation}
\scriptsize
\begin{aligned}
& \text{Select relay $k$ for reception where $k=\arg \max_{i, \left(Q(i)+\mathcal{R}_{ij}^{SR}\left(0\right)t\right)<Q_{max}}
\mathcal{R}_{ij}^{SR}\left(0\right)$} & & \max_{i, \left(Q(i)+\mathcal{R}_{ij}^{SR}\left(0\right)t\right)<Q_{max}} \mathcal{R}_{ij}^{SR}\left(0\right)   >
\max_{j,\mathcal{R}_j^{RD}\left(P_{max}\right)t<Q_{j} }\mathcal{R}_j^{RD}\left(P_{max}\right) \\
&\text{Select relay $k$ for transmission where $k=\arg \max_{j,\mathcal{R}_j^{RD}\left(P_{max}\right)t<Q_{j} }\mathcal{R}_j^{RD}\left(P_{max}\right)$
}
& & \max_{i, \left(Q(i)+\mathcal{R}_{ij}^{SR}\left(0\right)t\right)<Q_{max}} \mathcal{R}_{ij}^{SR}\left(0\right)   <
\max_{j,\mathcal{R}_j^{RD}\left(P_{max}\right)t<Q_{j} }\mathcal{R}_j^{RD}\left(P_{max}\right) \\\\
\label{eq:HDmodeexpression}
\end{aligned}
\end{equation}
}
\comments{
 $\arg \max \left\{ \max_{i, \left(Q(i)+\mathcal{R}^{ij}_{SR}\left(0\right)t\right)<Q_{max}}\left\{\frac{ P_s |g_i|^2}{N_0}\right\},
\max_{j,\mathcal{R}^j_{RD}\left(P_{max}\right)t<Q_{j} }\left\{\frac{ P_{max} |h_j|^2 }{N_0}\right\} \right\}$

}
\item If $\mathfrak{F_{R}}\neq\phi$, calculate throughput for all feasible links,
$\uptau_{i,j}\left(P_{j}\right)\,\forall\,\left\{\{i,j\}\in\mathfrak{F_{R}}\, \&\, P_j \in\left\{P_j^{min},P_{i,j}^{max}\right\} \right\}$.
Select the link pair and relay power, $P_j$, which yields the maximum $\uptau_{i,j}(P_j)$ as per \eqref{eq:optimalsolution}.
\comments{ $\left\{
\hat{P_{j}},\hat{i},\hat{j}\right\} =\arg\max_{P_{j},i,j}\left\{ \uptau\left(P_{j},i,j\right)\right\} $}
\end{enumerate}
}
\comments{
{\color{blue} Similar to a brute-force algorithm, JPASS calculates the optimal power along with the sum-throughput for all possible links, and then selects the links with highest sum-throughput. When $Q_{max} \rightarrow \infty$ and $Q(i)>>0\,\forall\, i$, i.e. the relays have infinite buffer and sufficient data available for transmission, the last two constraints of \eqref{eq:sum_throughput} are always satisfied. As a result, the buffer state does not impact the relay selection and power allocation and in this scenario, the JPASS algorithm is an optimal scheme and achieves the highest sum-throughput and serves as a benchmark. However, in the case of finite buffer, this scheme is not optimal. An optimal algorithm must utilize the information regarding buffer state transition over time to yield the maximum sum-throughput.}
}

\subsection{\bf Ratio Selection Scheme}

The JPASS algorithm calculates the optimal power along with the sum-throughput for all possible links, and then selects the links with highest sum-throughput. For example, if there are $n$ relays in the network, there can be at most $n(n-1)$ feasible links and thus, $n(n-1)$ sum-throughput values have to be calculated. So, in addition to JPASS, we propose a low computation, ratio selection scheme (RSS), in which the throughput is calculated for only one link. In RSS, the link pair is selected first and then the throughput is calculated only for the selected link pair. The selection criteria is proposed below.

In order to achieve high throughput and a good QoS, it is desirable to have a good S-R channel, $g_i$ as well as a good R-D channel, $h_i$,  and have lower interference, $e_{i,j}$ for $i,j \in {1,...,n} \,\&\, i \ne j$., which implies a poor inter-relay link. Based on this requirement, we come up with a heuristic ratio $\frac{g_i h_j}{e_{ij}}$ and select the link pair for which this ratio is maximum. This selection criteria can be mathematically expressed as
\begin{equation}
\text{Selected Link Pair } (\hat{i},\hat{j}) = \arg \max_{i \neq j}  \left( \frac{g_i h_j}{e_{ij}}\right) 
\end{equation}
It can be noted that this heuristic selection scheme will select the relay pair with a good S-R and R-D link and a poor interference link. As the relay selection is based on a ratio, this scheme is termed as ratio selection scheme (RSS). The main steps for RSS are as follows:
{
\begin{enumerate}
\item Find the set of feasible relays $\mathfrak{F_{R}}$ for which $P_j^{min}<P_{i,j}^{max}$.
\item If $\mathfrak{F_{R}}=\phi$, i.e. no relay pair satisfies the constraints in \eqref{eq:sum_throughput2}, follow step 2 of JPASS.
\item If $\mathfrak{F_{R}}\neq\phi$, select relay pair $\left\{ \hat{i},\hat{j}\right\} =\arg \max_{i \neq j}  \left( \frac{g_i h_j}{e_{ij}}\right)
$
\item Evaluate $\uptau_{\hat{i},\hat{j}}\left(P_{\hat{j}}\right)$ at extreme values of $P_{\hat{j}}$ i.e. $P_{\hat{j}}\in\left\{
P_{\hat{j}}^{min},\,P_{\hat{i},\hat{j}}^{max} \right\}$. 
Select value of relay power, $P_{\hat{j}}$ which yields the maximum throughput.
\comments{ $\left\{
\hat{P_{j}},\hat{i},\hat{j}\right\} =\arg\max_{P_{j},i,j}\left\{ \uptau\left(P_{j},i,j\right)\right\} $}
\end{enumerate}
}
\comments{ 

\begin{algorithm}
  \caption{Brute Force Selection Scheme (BFSS)}
  \label{OSSalgo}
  \begin{enumerate}
   
    \item Initialize $P_s$ and $P_j\,\, \forall \, j$.

    \item $\forall i \neq j$ obtain $P_j$ to maximize $\uptau(i,j)$. 

    \item If $\gamma_{ij} < \phi_{1}$ or $\gamma_{j}  < \phi_{2}$ then $\uptau(i,j)=0$.
      
	\item Selected relays are $(\hat{i},\hat{j})=\arg \max_{i,j} \uptau(i,j)$
  
  \end{enumerate}
\end{algorithm}

\begin{algorithm}
  \caption{Ratio Selection Scheme (RSS)}
  \label{RSSalgo}
  \begin{enumerate}
  
     \item Select relays $(\hat{i},\hat{j}) = \arg \max_{i \neq j}  \left( \frac{g_i h_j}{e_{ij}}\right) $.

	 \item Obtain $P_{\hat{j}}$ to maximize $\uptau(\hat{i},\hat{j})$ subject to $\gamma_{\hat{i}\hat{j}} \geq \phi_{1}$ and $\gamma_{\hat{j}}  \geq
	 \phi_{2}$.
	  
    \end{enumerate}
\end{algorithm}
}
\noindent \emph{\bf Complexity Analysis:} Both JPASS and RSS require complete CSI. The complexity of the proposed algorithms, quantified in terms of number of floating point operations (FLOPs) and the number of comparisons required, is given in Table 1. The complexity of HD relaying mode is ignored as it is the same for both schemes.  
\comments{	
	Overall number of FLOPs required for JPASS is $52n(n-1)+2$ and the number of comparisons required is $8n(n-1)$. Overall number of FLOPs required for RSS is $27n(n-1)+30$ and the number of comparisons required is $7n(n-1)+1$. 
}

{\small	
	\begin{center}
		\begin{tabular}{c c c } 
			\hline
			  & FLOPs & Comparisons \\
			\hline\hline
			JPASS & $52n(n-1)+2$ & $8n(n-1)$  \\ 
			\hline
			RSS & $27n(n-1)+30$ & $7n(n-1)+1$  \\
			\hline
		\end{tabular}
	\end{center}

	\comments{
	The first two steps of both the algorithms are same and hence have the same complexity. 
	
	As mentioned previously JPASS requires that, the sum throughput is calculated for all possible $n(n-1)$ possible links ($2n(n-1)$ throughput calculations). This requires that the 2 possible relay powers are calculated for all possible links ($2n(n-1)$ relay power calculations). The link and power with the highest throughput is selected ($2n(n-1)$ comparisons???).  
	
	On the contrary, RSS calculates the ratio based metric for all possible links ($n(n-1)$ calculations). The relays corresponding to the highest metric are selected ($n(n-1)$ comparisons???). The sum-throughput of the selected relays is evaluated using the minimum and maximum power values and the best is chosen (2 throughput calculation and 1 comparison).  
	
	So we have, JPASS ($4n(n-1)$ calculations and $2n(n-1)$ comparisons), RSS ($n(n-1)+2$ calculations and $n(n-1)+1$ comparisons).
 	}
}

\noindent \emph{\bf Remark:}
{
In order to calculate the analytical expression of the average sum-throughput, $\mathbb{E}\left[\uptau_{i,j}(P_j)\right]$, the distribution of $P_j^{min} \leq P_{j} \leq  P_{i,j}^{max}$ is required for all possible $i \neq j$. It can be noted that, $P_j^{min}=\max \left\{
	\frac{\phi_{2}N_{0}}{\left|h_{j}\right|^{2}}, \frac{P_s|g_i|^2 }{\left(2^{\left(Q_{max}/t-Q_i/t\right)}-1\right)|e_{ij}|^2 }-\frac{N_0 }{|e_{ij}|^2 } \right\}$ and $
	P_{i,j}^{max}=\min\left\{P_{max},\frac{P_{s}\left|g_{i}\right|^{2}-\phi_{1}N_{0}}{\phi_{1}\left|e_{ij}\right|^{2}},
	\frac{N_0\left(2^{Q_j/t}-1\right)}{|h_j|^2}\right\}$ are correlated random variables (RVs) due to the reciprocity assumption and also because all possible combinations of $i \neq j$ must be considered. For example, consider a case of $n=3$ relays; for $j=1$ and $j=2$, $P_{i,1}^{max}$ and $P_{i,2}^{max}$, both are functions of the RVs $g_3$ and $e_{21}=e_{12}$ and thus, are correlated RVs. Therefore, in this case, and also for the general $n$ case, it gets intractable to derive the theoretical expression for the average sum-throughput. 
}

\section{Numerical Simulations}


{ Monte-Carlo simulations were performed in MATLAB to analyze and compare the average sum-throughput of the proposed selection schemes. $5\times10^4$ realizations of the channel power gains (exponential random variables) were generated for each value of $P_{max}$ and $n$. The sum-throughput
for all the schemes was calculated and averaged to yield the mean sum-throughput. The sum-throughput of the proposed JPASS and RSS is compared with the
existing selection schemes in the literature, e.g., max-max relay selection (MMRS) \cite{BibIkhlef2012}, BA-SOR selection scheme proposed in
\cite{BibNomikos2014} and the min-power scheme proposed in \cite{BibNomikos2015}. In addition, the average sum-throughput of the conventional half-duplex (CHD)
relaying scheme is also plotted. Without loss of generality, the mean channel powers and the mean noise power are assumed to be unity. In the
simulations, all relays are assumed to have a buffer of size, $Q_{max}$, and initially, the amount of data stored inside the buffer of all relays is denoted by $Q_{s}$. $Q_{max}=20^8$ indicates the scenario of infinite buffer size and $Q_{s}=0$ indicates an empty buffer.

\begin{figure}             
	\centering
	\includegraphics[width=0.9\columnwidth]{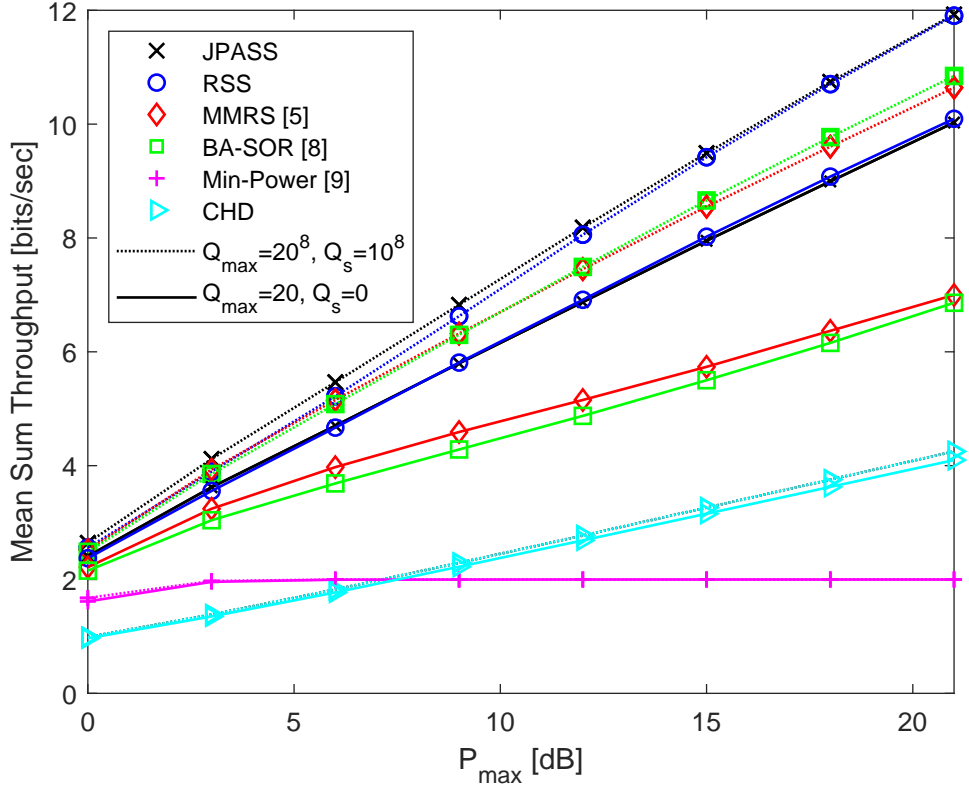}
	\caption{Average sum-throughput achieved by the proposed schemes where $n=6$, $R_1=R_2=1$ and $P_s=P_{max}$.}
	\label{fig1}
\end{figure}

\figref{fig1} shows the average sum-throughput achieved by the selection schemes varying the maximum transmit power, $P_{max}$, where $R_1=R_2=1$ and $P_s=P_{max}$. It can be observed that for all schemes, the mean sum-throughput increases with increase in $P_{max}$. For the infinite buffer case, it can be observed that JPASS yields the maximum sum-throughput. The RSS closely follows and achieves the sum-throughput equivalent to JPASS when the nodes can transmit with higher power. For $P_{max}=21$dB, JPASS and RSS yield approximately 20\% higher throughput compared to that of MMRS and BA-SOR. The sum-throughput of the min-power scheme is limited to two because the nodes transmit with the minimum power which satisfies the rate requirement i.e. $R_1=R_2=1$. For the finite buffer case, the sum-throughput of all the schemes is lower because the number of links which maybe activated for transmission is lower due to either full or empty buffers at the relays. However, the gain in sum-throughput, at $P_{max}=21$dB, of the proposed JPASS and RSS is approximately 40\% higher compared to MMRS and BA-SOR. In both MMRS and  BA-SOR, the relays always transmit with maximum power, $P_r=P_{max}$, due to which the S-R link throughput reduces because of high interference. As a result, the number of feasible links (which satisfy the minimum throughput requirement along with the buffer constraints) also reduces. The proposed schemes, on the contrary, reduce the transmit power of the relay, causing reduced interference and increasing the number of relays in the feasible set. Both these factors contribute towards the gain in sum-throughput. Moreover, it can be noted that all schemes mimicking full duplex relaying give significant increase in sum-throughput compared to CHD.

\figref{fig2} shows the average sum-throughput achieved by varying the number of relays. The rate requirement in this case is higher i.e. $R_1=R_2=3$ and $P_s=P_{max}=15$dB. For all the schemes, the sum-throughput increases with the increase in the number of relays $n$. Again it can be observed, that for the infinite buffer scenario, JPASS yields the highest sum-throughput and the RSS closely follows. MMRS and BA-SOR yield a lower sum-throughput compared to both JPASS and RSS. The sum-throughput achieved by min-power scheme is higher in this case because minimum rate requirement is three times higher compared to \figref{fig1}. However, similar to \figref{fig1}, the sum-throughput saturates to a maximum value due to the reason discussed previously. For the finite buffer case, the sum-throughput of all the schemes is lower compared to the infinite buffer case. Even as the number of relays increases, the proposed schemes offer 5-10\% gain in sum-throughput.   
}
  
 \begin{figure}             
\centering
  \includegraphics[width=0.9\columnwidth]{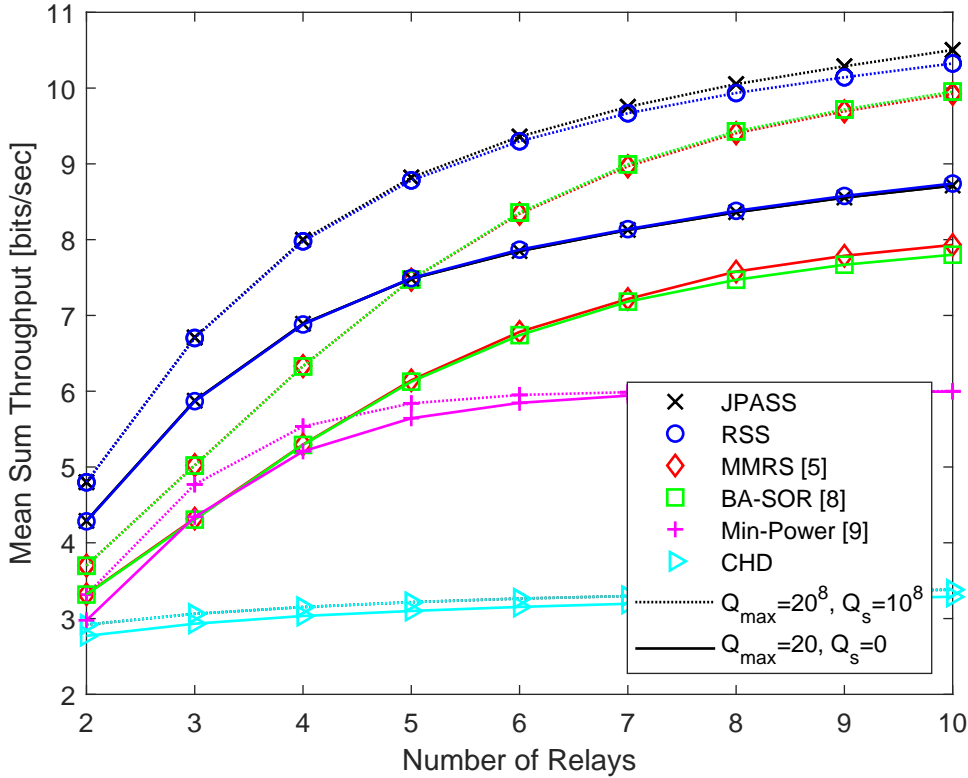}
	\vspace{-8pt}
  \caption{Average sum-throughput achieved by the proposed schemes where $R_1=R_2=3$ and $P_s=P_{max}=15$dB.}
  \label{fig2}
\end{figure}
 
\comments{
\figref{fig3} shows the average sum-throughput achieved by the selection schemes where $R_1=R_2=1$ and the source power is assumed to be
significantly higher than the maximum relay power i.e. $P_s = 10 P_{max}$. Again, it can be observed that the optimal scheme, OSS, yields the maximum
throughput. Similar trends in throughput can be observed for all the schemes. However, it can be noted that RSS gives higher throughput as compared to
the previously proposed MMRS and MMT schemes in all scenarios. 

\begin{figure}             
\centering
  \includegraphics[width=0.9\columnwidth]{PsPmaxK3K12Ps10Pmax.eps}
  \caption{Average sum-throughput achieved by the proposed schemes where $R_1=R_2=1$ and $P_s=10 P_{max}$ dB.}
  \label{fig3}
\end{figure}
}

{
	The end-to-end throughput for a buffer-aided cooperative relaying system is dominated by the weakest hop throughput, therefore another important performance metric is minimum of the average throughput of each hop \cite[Eq. (6)]{BibIkhlef2012}. \figref{fig3} plots the minimum of the mean hop-throughput (MMHT) for the proposed schemes. \figref{fig3} shows that both JPASS and RSS outperform the existing schemes and have a higher MMHT. It can be observed that the MMHT for the infinite buffer case is lower than that of the finite buffer case. This is because when the relays have large buffers filled with data, the proposed algorithms try to select a higher relay transmit power to increase the R-D transmission rate and eventually to maximize the sum-throughput. This affects the S-R link throughput due to the higher inter-relay interference. As a result, the MMHT is dominated by the low S-R hop-throughput. On the contrary, when the buffer size is small, the buffer may not have a lot of data stored, which limits the relay power to a lower value (the relay cannot transmit at a rate higher than the data stored in its buffer). As a result, the interference to the the S-R link is lowered and the MMHT is not affected the way it got in the prior case.  
}

\begin{figure}             
	\centering
	\includegraphics[width=0.9\columnwidth]{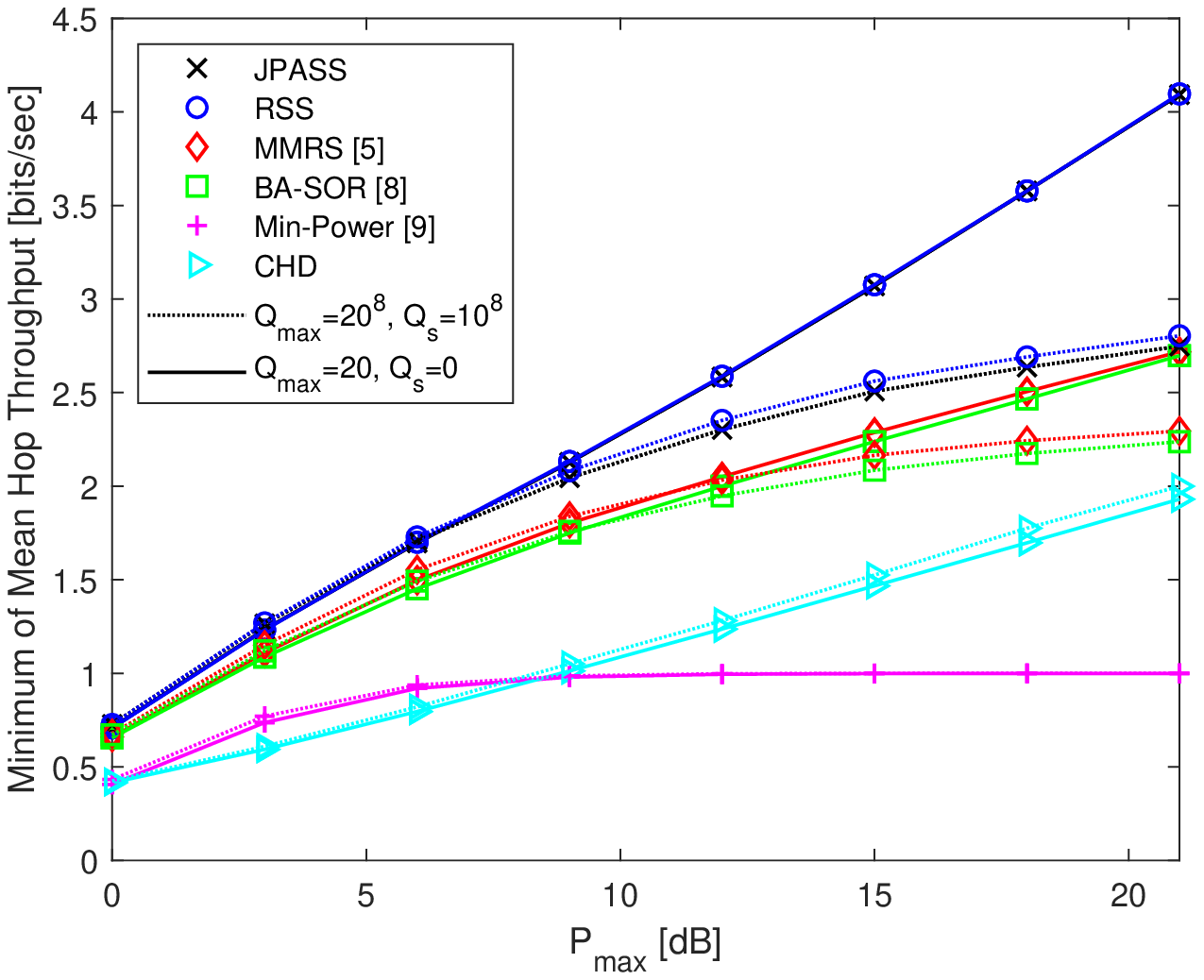}
	\caption{Minimum of average throughput of the S-R hop and the R-D hop, where $n=3$, $R_1=R_2=1$ and $P_s=P_{max}$.}
	\label{fig3}
\end{figure}

\begin{figure}             
	\centering
	\includegraphics[width=0.9\columnwidth]{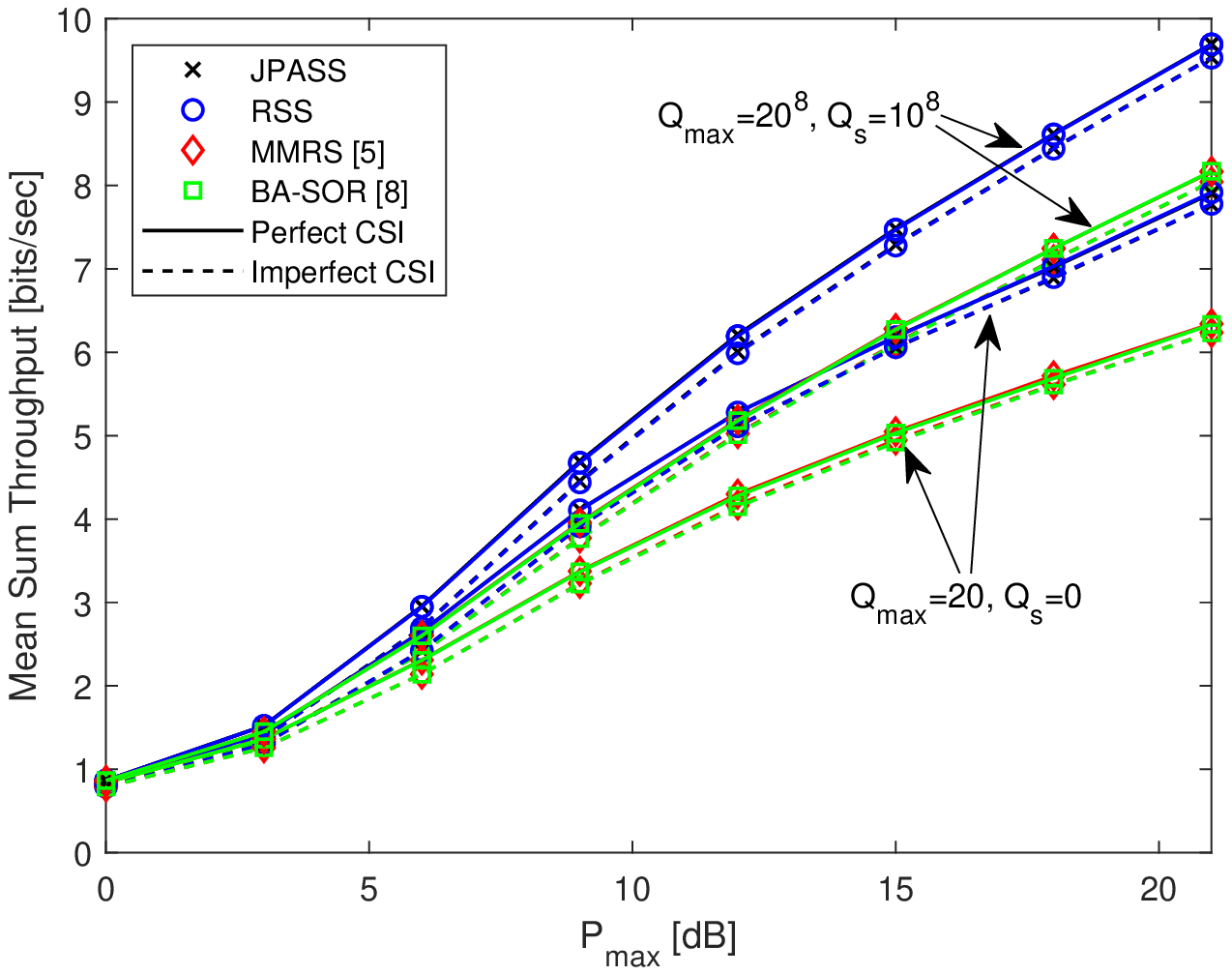}
	\caption{Average sum-throughput achieved by the proposed schemes under imperfect CSI, where $n=3$, $\sigma_{\eta}^2=0.1$, $R_1=R_2=2$ and $P_s=P_{max}$.}
	\label{fig4}
\end{figure}

{    
	In order to evaluate the performance of the proposed schemes under imperfect channel state information (CSI), we follow the model in \cite{BibAmin2011,BibSeyfi2012}, where the exact channel is related to the estimated channel as $h=\hat{h}+\eta$, such that $h$ denotes the perfect channel, $\hat{h}$ denotes the channel estimate and $\eta$ stands for the estimation error. When the channel is zero mean and an unbiased estimator is designed, the variance of the estimation error, $\sigma_{\eta}^2$, is related as $\sigma_h^2=\sigma_{\hat{h}}^2+\sigma_{\eta}^2$ \cite{BibAmin2011,BibSeyfi2012}. \figref{fig4} shows the performance of the proposed schemes in case of imperfect CSI where the estimation error variance for the S-R channel, R-D channel and the inter-relay channel is assumed to be the same i.e. $\sigma_{\eta}^2=0.1$. It can be observed that there is degradation in the achievable sum-throughput when perfect CSI is not available at the transmitter. Moreover, it can be shown that as $\sigma_{\eta}^2$ increases, the achievable sum-throughput decreases. 
}

\section{Conclusion}

In a two-hop buffered relay network, joint relay selection and transmitting-relay's power optimization can significantly
enhance the sum-throughput.\comments{The power of the transmitting-relay is an important parameter that significantly impacts the sum-throughput of
this network.} Two relay selection schemes namely; 1) Joint relay power allocation and selection scheme (JPASS) and 2)  Ratio selection scheme (RSS),
have been proposed. Numerical simulations show that both the proposed schemes significantly enhance the sum-throughput compared to existing schemes.

\bibliography{IEEEfull,document}
\bibliographystyle{IEEEtran}


\clearpage

\newpage

\end{document}